\documentclass[amsmath, amsfonts, amssymb, twocolumn,superscriptaddress,pra]{revtex4-1}
\usepackage{graphicx}
\usepackage{color}
\usepackage[normalem]{ulem}
\usepackage{ulem}
\DeclareMathOperator{\sinc}{sinc}
\DeclareMathOperator{\tanc}{tanc}

\begin{document}

\title{Analytic instability thresholds in folded Kerr resonators of arbitrary finesse}

\author{William J.~Firth\footnote{To whom correspondence should be addressed.}\email[]{w.j.firth@strath.ac.uk}}
\affiliation{SUPA and Department of Physics, University of
Strathclyde, 107 Rottenrow East, Glasgow G4 0NG, UK}

\author{John B. Geddes}
\affiliation{Olin College of Engineering, Needham MA 02492}

\author{Nathaniel J. Karst}
\affiliation{Babson College, Babson Park MA 02457}

\author{Gian-Luca Oppo}
\affiliation{SUPA and Department of Physics, University of
Strathclyde, 107 Rottenrow East, Glasgow G4 0NG, UK}

\date{\today}

\begin{abstract}
We present analytic threshold formulae applicable to both dispersive (time-domain) and diffractive (pattern-forming) instabilities in Fabry-Perot Kerr cavities of arbitrary finesse. We do so by extending the gain-circle technique, recently developed for counter-propagating fields in single-mirror-feedback systems, to allow for an input mirror. In time-domain counter-propagating systems walk-off effects are known to suppress cross-phase modulation contributions to dispersive instabilities. Applying the gain-circle approach with appropriately-adjusted cross-phase couplings extends previous results to arbitrary finesse, beyond mean-field approximations, and describes Ikeda instabilities.
\end{abstract}

\maketitle
\section{Introduction}
Diffractive optical pattern formation in driven nonlinear media has been studied extensively since the 1980s, especially in ring resonators.  While spontaneous pattern formation is particularly rich in two transverse dimensions, there is an important and close analogy between diffraction in one transverse dimension and dispersion in the time domain, with sideband instabilities leading to spontaneous oscillations analogous to one-dimensional patterns, both describable in the high-finesse limit by the Lugiato-Lefever equation (LLE) \cite{LLE87}. Dispersive instabilities have a close connection to the generation of frequency combs and cavity solitons, topics of enormous current interest and importance \cite{Pasquazi18}.

An important group of pattern-forming systems  are double-pass schemes, including counter-propagating (CP) beam configurations with two input beams \cite{grynberg93, Geddes1994}, single feedback mirror (SFM) configurations with a single input beam \cite{firth90a}, and Fabry-Perot (FP) resonators (Fig. \ref{cavityfig}), which is our main topic of investigation. 

A simple, but very powerful and general, technique to obtain thresholds for pattern formation in  SFM  systems has been recently proposed and demonstrated \cite{Firth2017}. It basically shows that the ratio $f/b$ of the relative perturbations of the forward- and backward-traveling fields always lies on a circle as the input phase of $b$ is varied, even in highly-lossy nonlinear media. It is then a matter of simple geometry to identify conditions under which this $f/b$ gain-circle allows the instability threshold condition (e.g. $f=b$ at the mirror) to be met. Here we develop and extend this technique, demonstrating its applicability to cavities and to dispersive instabilities. Previous results in this area have limited scope and applicability, but the gain circle method is fully general, spanning existing models and unifying previous results.  In particular, it applies to the microresonator systems which are the leading technology in frequency comb and soliton generation \cite{Pasquazi18}. Important potential applications include the opto-mechanical control of Bose-Einstein condensates in optical cavities of lower \cite{Brennecke08} and higher \cite{Wolke12} Purcell factors as well as polariton micro-cavities \cite{Sich11}.
\section{Fabry-Perot Cavity Model.} 
\begin{figure}
\centering
\includegraphics[width=\columnwidth]{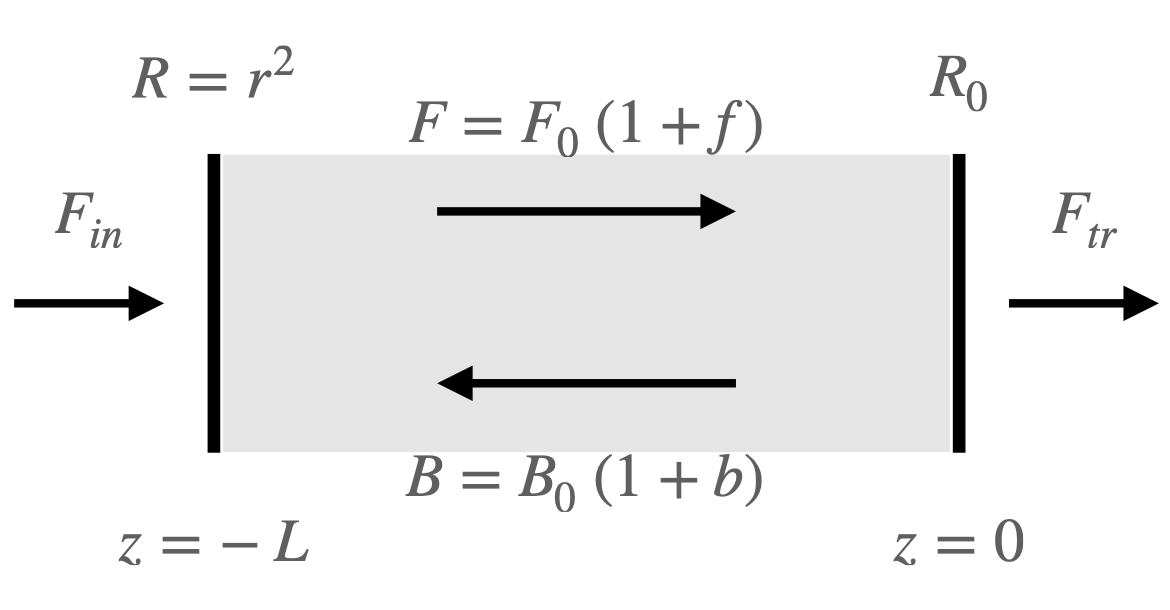}
\caption{A FP cavity of length $L$, filled with a nonlinear Kerr medium. A uniform pump field $F_{in}$ enters at $z=-L$ through a mirror of reflectivity $r$ and drives forward ($F$) and backward ($B$) fields in the cavity, establishing zero-order cavity fields $F_0$ and $B_0$, whose stability depends on relative perturbations $f$ and $b$. A transmitted field $F_{tr}$ exits the cavity at $z=0$.}
\label{cavityfig}
\end{figure}

%
We start by considering  the  case of a FP cavity with a Kerr nonlinearity, as shown in Fig. \ref{cavityfig}. 
The mirror reflectivities are assumed real, any reflection phase being subsumed into the linear cavity phase $\phi_0$. The evolution equations for the forward, $F$, and backward, $B$, fields in the Kerr medium are given by
\begin{subequations}
\label{KCFPeqs}
\begin{align}
\frac{\partial F}{\partial z} +\beta_1 \frac{\partial F}{\partial t} & = -i \hat{D} F +i \left( |F|^2+G|B|^2 \right) F, \\
-\frac{\partial B}{\partial z}+\beta_1 \frac{\partial B}{\partial t} & = -i \hat{D} B +i \left( G|F|^2+|B|^2 \right) B.
\end{align}
\end{subequations}
Cross-phase modulation (XPM), i.e. the extent to which the standing-wave modulation of the cavity field generates a corresponding modulation in the nonlinear index, is described by the grating-parameter $G$, as in \cite{Geddes1994}. To allow for propagation effects, we  include first order dispersion ($\beta_1 = v^{-1}_{g}$, where $v_g$ is the group velocity). In general,  the operator $\hat{D}$  is given by $(\beta_2 / 2) (\partial^2 / \partial t^2)-\nabla^2_\perp /2k$, where $\beta_2$ is the GVD coefficient and $t$ is the (fast) time in (\ref{KCFPeqs}), while the operator  $ \nabla^2_\perp /2k$, acts on the transverse coordinate(s) with $k$ being the light wavenumber. The governing equations are supplemented with the appropriate boundary conditions at the left ($z=-L$) and right ($z=0$) mirrors
\begin{subequations}
\label{KCFPeqsb}
\begin{align}
F(-L) & = \sqrt{1-r^2}F_{in} + r e^{i\phi_0} B(-L), \\
B(0) & = \sqrt{R_0}F(0), 
\end{align}
\end{subequations}
where $F_{in}$ is the input field, $r^2$ and $R_0$ are the reflectivities of the input and back mirrors. Solving (\ref{KCFPeqs}) subject to (\ref{KCFPeqsb}) at zero order, with $F=F_0(z)$, $B=B_0(z)$, we find that both $|F_0|^2=I$ and $|B_0|^2=R_0I$ are conserved through the medium, resulting in:
\begin{equation}
\label{zeroorder}
F_0(-L) = \sqrt{1-r^2}F_{in} + r \sqrt{R_0} e^{i(\phi_0+\phi_{nl})} F_0(-L)
\end{equation}
The nonlinear phase $\phi_{nl} = (1+R_0)(1+G)IL$. 

We examine the stability of the homogeneous zero-order solution by linearizing (\ref{KCFPeqs})-(\ref{KCFPeqsb}), and seeking a non-trivial first order solution. Using the ansatz $F = F_0(1+f(z,t))$, with a similar form for $B$, we obtain the linearized evolution equations,
\begin{eqnarray}
\label{KCFPdisp}
\frac{\partial f}{\partial z} + \beta_1 L \frac{\partial f}{\partial t} &=& -i \theta f  + iIL\left(f+f^*+GR_0(b+b^*)\right), \nonumber \\
\\
-\frac{\partial b}{\partial z} + \beta_1 L \frac{\partial b}{\partial t} &=& -i \theta b + iIL\left(G(f+f^*)+R_0(b+b^*)\right), \nonumber
\end{eqnarray}
where we have scaled $z$ by $L$. Although the evolution equations (\ref{KCFPdisp}) are general, the dimensionless parameter $\theta$ depends on what type of instability we are considering. For example, dispersive instabilities with frequency $\pm \Omega$ give $\theta = -\beta_2 \Omega^2 L/2$. Physically, $\theta$ is then a dimensionless measure of the dispersive phase mismatch between the sidebands and the zero-order field over distance $L$.

These propagation equations are essentially identical to those in the CP analysis of \cite{Geddes1994}, which suggests a unified approach to all configurations of Kerr media. We emphasize that (\ref{KCFPdisp}) are local equations, holding throughout any dispersive or diffractive Kerr medium independent of any particular boundary conditions. 
 
The  boundary conditions for $(f,b)$ follow from linearization of (\ref{KCFPeqsb}) and use of the zero-order solution (\ref{zeroorder}):
\begin{eqnarray}
\label{KCFPbc}
f(-1) =  r e^{i(\phi_0+\phi_{nl})} b(-1), \;\;\;\;\;\;\;\;\
f(0) =  b(0) 
\end{eqnarray}
The usual nonlinear and dispersion length scales can be recognized in (\ref{KCFPdisp}), but also a third length scale  $|\beta_1\Omega|^{-1}$, the walk-off length \cite{Yu98}. It reflects the fact that the CP field encountered in a round trip of the cavity is up to a round-trip  ahead or behind in time. 
 
For any finite $\Omega$, the XPM cross-coupling of the CP fields will suffer a phase-mismatch on walk-off length scale. This complicates the analysis of the dispersive case, and so it will be convenient to postpone detailed discussion until we have solved the diffractive problem. We will now use the gain-circle technique to develop analytic solutions for the diffractive case, and will then show that the dispersive problem can be well approximated within the same framework.
\section{Gain-circle Model for Diffractive Instability}
 %
The walk-off problem does not arise for pure transverse instabilities, which are zero-frequency. The perturbation equations (\ref{KCFPdisp}) then simplify, because we can ignore the time derivative terms. With $\hat{D} $ as a purely diffraction operator, $\theta =  K^2L/2k$ now characterizes the transverse wavevector $K$. Physically $\theta$ corresponds to the diffractive phase shift between the sidebands and the zero-order field over the length $L$ of the cavity, analogous to the dispersive case.

After dropping the time-derivatives, the resulting propagation equations can be solved exactly and have been extensively analyzed in relation to CP and SFM pattern formation, but not for two-mirror (resonator) problems. In SFM problems there is no input perturbation and so $f(0)=0$, meaning that $f/b$ is a measure of gain. The ``gain" $f/b$ at the mirror traces out a circle as the phase of $b$ at the input plane is varied \cite{Firth2017}. We now extend this ``gain circle" approach to the solution of the FP problem, which requires cavity boundary conditions. Step by step calculations are presented in the Appendix.

Threshold formulae for any such problem can be found with the gain circle method. Starting from one mirror, with the phase of $b$ arbitrary, we integrate the propagation equations and determine the parameters of the gain circle at the other mirror. An instability occurs if any point on the gain circle satisfies the boundary condition there. This procedure works for any Kerr cavity, and can also cater for losses and other complications (see \cite{Firth2017} and  Appendix), using numerical methods in general. Here we concentrate on the particular case $R_0=1$ of the FP cavity in Fig.~\ref{cavityfig}. As we will show, it also allows exact  solution, giving analytic threshold formulae. While both directions of integration lead to the same analytic formula, backward integration is neater, and so we present that approach here.

\begin{figure}
\centering
\includegraphics[width=0.9\columnwidth]{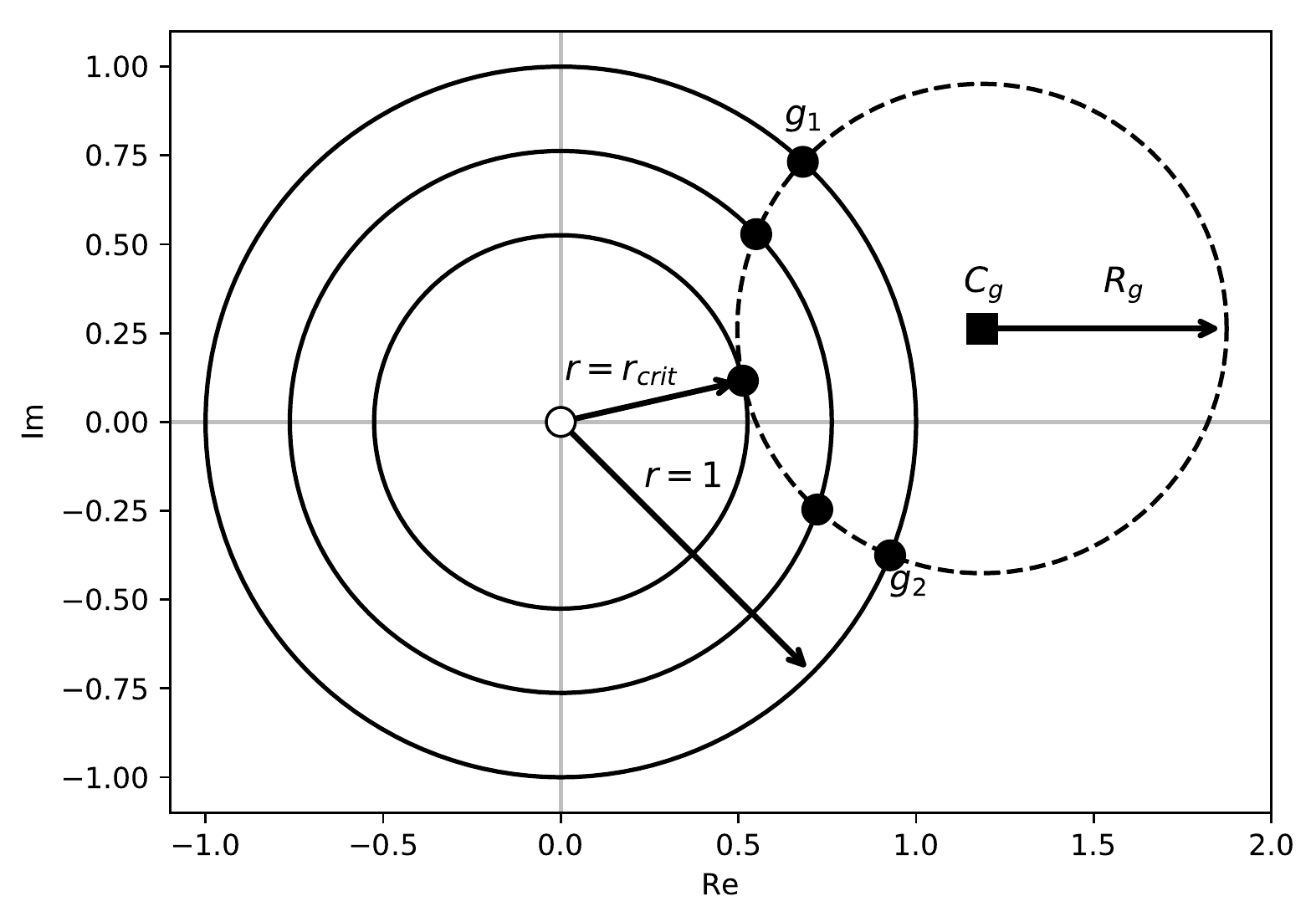}
\caption{Unit circle (thick) centred on the origin of the complex plane. A gain circle (dashed) intersects the unit circle at points ($g_1,g_2$) given by (\ref{KCFPanal}). The centre $C_g$ and radius $R_g$ of the gain circle  are given by (\ref{RgCg}) for a FP Kerr cavity. Intersection of the gain circle and any ``loss circle" of radius $r$, $1 > r > r_{crit}$ enables threshold condition  (\ref{KCFPthresh}) to be satisfied for two different values of $\phi$ at that reflectivity $r$. At  $r = r_{crit}$ the gain and loss circles touch, marking the limit of instability. $(\theta, IL) = (0.4, 0.1)$.}
\label{FPcircleplot}
\end{figure}

Dropping the time derivatives in (\ref{KCFPdisp}), and setting $R_0=1$, $b=-f^*$ and $b=f^*$ are self-consistent special ``modes". Labelling them $f_1, f_2$ respectively, we find:
\begin{subequations}
\label{KCFPeigen}
\begin{align}
\frac{d^2f_n}{dz^2} & = - \psi_n^2 f_n, \; n=1,2 \\
\psi_n^2 & = \theta \left( \theta -2 \left( 1+(-)^n G \right) IL \right).
\end{align}
\end{subequations}
The key parameters $\psi_1, \psi_2$  are precisely the parameters defined in the stability analysis of the diffractive CP problem for a Kerr medium \cite{Geddes1994}. Because the  $\psi_n^2 $ are real, $f_n^*$ obeys the same differential equation, enabling construction of two linearly independent solutions of the system (\ref{KCFPdisp}), and hence of gain circles. The boundary conditions at the $z=0$ mirror is $f=b$, but the phase of $f$ and $b$ is not fixed, and we can match on to modes 1 and 2  with the choices $f_1(0)=b_1(0)=i$ and $f_2(0)=b_2(0)=1$. Solving, we readily obtain for the resultant gains $g_n = f_n(-1)/b_n(-1)$: 
\begin{subequations} 
\label{KCFPanal}
\begin{align}
g_1 &= \frac{cos\psi_1+i\theta \sin\psi_1/\psi_1}{cos\psi_1-i\theta \sin\psi_1/\psi_1}=e^{i\phi_1}, \\
g_2 &= \frac{ \cos\psi_2 + i \psi_2\sin \psi_2 /\theta}{\cos\psi_2 -i \psi_2\sin \psi_2 /\theta}=e^{i\phi_2}.
\end{align}
\end{subequations}
The real and imaginary terms  are explicit  in (\ref{KCFPanal}), because all expressions are even in $\psi_n$, and so are real. Because both modes obey $|f|=|b|$ by definition, $|g_n|=1$ and so the $\phi_n$ are real. As in \cite{Firth2017}, $g_1$ and $g_2$ define a gain circle. Its radius $R_g$ and centre $C_g$ are given by
\begin{equation}
\label{RgCg}
 R_g= \left| \frac{e^{i\phi_1}-e^{i\phi_2}}{e^{i\phi_1}+e^{i\phi_2}} \right |, \;\;\;\;\;\;\; C_g = \frac{2}{e^{-i\phi_1}+e^{-i\phi_2}} \, . 
\end{equation}
Note that $C_g$ is outside the unit circle (unless $g_1=g_2$, when $R_g =0$). Since both $g_1$ and $g_2$ lie on the unit circle, the gain-circle arc between them lies {\em{inside}} the unit circle, and so can pass through the point $g=re^{i\phi}$, fulfilling the boundary condition  (\ref{KCFPbc}). Fig. \ref{FPcircleplot} illustrates the geometry of the gain circle intersecting the unit circle and smaller ``loss-circles" of various radii $r$.

At instability threshold the distance from $re^{i\phi}$ to the centre of the gain circle must equal its radius $R_g$. Imposing this condition leads to the desired formula:
\begin{equation}
\label{KCFPthresh}
\frac{1+r^2}{2r} = Re(C_g e^{-i\phi})
\end{equation}
This threshold formula is our key result. It  holds for all values of $r$ and $\phi$ and  matches all previous special case analyses. It is also valid for all values of the XPM parameter $G$ \cite{Hill20}, and thus (\ref{KCFPthresh}) holds for  Kerr liquids, or indeed any Kerr-like material, as well as for dielectrics where $G=2$.

For a given input reflectivity  $r$, cavity phase $\phi_0$, and input intensity $I_{in}$, $IL$ is known from the zero-order solution, and so (\ref{KCFPthresh}) is effectively an {\em{analytic}} formula from which the threshold values of $\theta$, and hence the transverse wavevector $K$, can be calculated. There are no solutions unless the gain circle intersects a circle of radius $r$ centred on the origin. The marginal case is when the latter circle touches the gain circle, see Fig. \ref{FPcircleplot}. This happens at a critical reflectivity $r_{crit}$ and total phase $\phi_{crit}$, which from simple geometry equals $(\phi_1 + \phi_2)/2$. Then  (\ref{KCFPthresh}) simplifies considerably, and one obtains 
\begin{equation} 
\label{KCFPcrit}
\frac{1+r_{crit}^2}{2 r_{crit}} = \frac{1}{|\cos((\phi_1-\phi_2)/2)|}
\end{equation}

The special case $\theta=0$, which corresponds to a plane-wave instability, generates the well-known criterion for vertical slope in the plot of $IL$ vs $I_{in}$ and the instability usually termed optical bistability (OB). In that limit, the analytic formula  (\ref{KCFPthresh}) yields:
\begin{equation} 
\label{KCFPswitch}
\frac{1+r^2}{2r} = \cos(\phi_0+\phi_{nl}) - \phi_{nl} \sin(\phi_0+\phi_{nl})
\end{equation}
This is the OB formula for arbitrary finesse. It reduces to a mean-field LLE-like formula when $1-r = \delta$ and the phases are small. Allowing $\theta$ to be finite, but also small, we obtain the mean-field approximation to (\ref{KCFPthresh}):
\begin{equation} 
\label{HFFP}
(\phi - 2\theta)(\phi - 2\psi_{2}^2/\theta) +\delta^2 = 0.
\end{equation}
Only if $\phi$ lies between $2\theta$ and $ 2\psi_{2}^2/\theta$ can this real equation have real roots. It is quadratic in any of the underlying parameters ($IL, \phi_0, \theta$). The transition from zero to two real roots of the equation  can be identified with contact, and then intersection, between the gain and loss circles. 

Far from the  mean-field limit, i.e. for $r \to 0$,  (\ref{KCFPthresh}) remains valid provided the radius $R_g$ of the gain circle diverges. Taking appropriate limits leads to the SFM threshold condition for Kerr media:
\begin{equation} 
\label{KCFPsfm}
\cos \psi_1 \cos\psi_2  + \left( \frac{\psi_2}{\psi_1} \right) \sin \psi_1 \sin \psi_2 = 0 \, .
\end{equation}
This formula matches the results of \cite{Firth2017} for an SFM system with no free-space section ($D=0$). It is actually identical to the {\em{even}} instability mode of the mirrorless CP instability \cite{Geddes1994}, for which $f=b$ (in our notation) holds in the centre of the medium. This points the way to further generalizations of our present analysis.
\section{Gain-circle Model for Dispersive instability}
%
A fuller discussion of the transverse problem would divert us from our other topic, which is dispersive instabilities. The gain circle method relies on $(f,b)$ being dependent on $z$ alone. For an oscillatory instability with frequencies $\pm \Omega$, retarded time transformations on $f$ and $b$ easily eliminate the time-derivatives from (\ref{KCFPdisp}), but at the expense of introducing explicit time dependence into the XPM terms (those prefixed by $G$). This is a manifestation of the walk-off associated with counter-propagation that was mentioned earlier. 

Close enough to zero GVD ($\beta_2 \sim 0$) the walk-off length dominates, enabling (\ref{KCFPdisp}) to be solved exactly. Firth analyzed this case for a number of resonator problems in \cite{firth81} with particular emphasis on our present configuration of a Kerr FP with $R_0=1$. He found threshold conditions for optical bistability (OB) and for side-mode (P1) and period-doubling (P2) instabilities of Ikeda type \cite{Ikeda79,Haelterman93}. Only P2 could occur on the positive-slope branches of the characteristic. The P1 thresholds occurred exclusively on the unstable negative-slope steady-state branches, because the walk-off effect reduces the effective nonlinearity compared to (zero-frequency) OB. Yu et al  \cite{Yu98, Yu98a} recovered and confirmed these results as a special case of a very general (and rather complex) analysis.

It is clear that walk-off means that the XPM contribution of the perturbations in (\ref{KCFPdisp})  will be weakened. Indeed Yu et al  \cite{Yu98} find that XPM is essentially negligible whenever the external mirrors provide the dominant coupling between the forward and backward intensities. This will usually be the case in a FP resonator.

This suggests setting $G=0$ in (\ref{KCFPdisp}), thereby eliminating XPM in these equations. Then the $f$ and $b$ equations decouple, and the $\beta_1$ terms can be removed from each by (separate)  phase transformations.  We can thus apply the same gain circle technique as for the transverse problem, but now with $\psi_1=\psi_2=\psi$, where $\psi^2= \theta(\theta -2IL)$. This degeneracy  does not make (\ref{KCFPanal}) trivial, and all the gain circle considerations still apply for $G=0$, as do the threshold formulae (\ref{KCFPthresh}), (\ref{KCFPcrit}), (\ref{HFFP}). However, because there is no walk-off at zero-order we retain $G=2$ in the nonlinear phase shift $\phi_{nl}$, which enters the boundary conditions (\ref{KCFPbc}) and the OB equation (\ref{KCFPswitch}). 

Recently Cole et al \cite{Cole2018} presented a mean-field model of LLE type for a dispersive Kerr FP, aimed at ultra-finesse microresonator systems. As with the LLE, both losses and pumping are distributed. The major difference from the ring cavity LLE is that there is an additional Kerr-like term proportional to the cavity-averaged intensity. This is directly comparable to equation (6) of Ref.~\cite{firth81} where the cavity-averaged term explicitly describes XPM due to counter-propagation.
  
We can directly compare the mean-field threshold formula of Ref.~\cite{Cole2018} with our expression (\ref{HFFP}) by setting $G=0$ in $\psi_2$ (only). There is perfect agreement, once the parameters of the two models are matched up using the vertical-slope condition (\ref{KCFPswitch}). For zero GVD we note that all finite-frequency instabilities occur on the negative-slope branch, as for the non-Ikeda instabilities in Ref.~\cite{firth81}. It follows that the sideband instabilities observed in the model of Ref.~\cite{Cole2018} are entirely due to finite $\beta_2$, as is clear from the fact that the frequency formulae of \cite{Cole2018} diverge as GVD approaches zero.

This adapted gain-circle model is not limited to high finesse or mean-field approximations. Used in (\ref{KCFPthresh}) it leads to a general dispersive instability formula
\begin{equation} 
\label{GFFP}
 \cos \phi \cos 2\psi - 2(IL-\theta)\sin \phi \sinc 2\psi= (-1)^n\frac{1+r^2}{2r} .
\end{equation}
This generalizes the results of \cite{firth81} to finite GVD: for zero GVD, $\theta =0$ and both $\cos(2\psi)$ and $\sinc(2\psi)$ go to unity, recovering Firth's results.

The factor $(-1)^n$ multiplying the right side arises because the boundary conditions require only that instabilities occur at frequencies which obey $\Omega t_R = n\pi$ with $n$ being an integer and $t_R$ the cavity round-trip time. For $n$ even, (\ref{GFFP}) extends LLE-type sideband instability expressions beyond the mean-field limit.  For $n$ odd, there is no mean-field limit to (\ref{GFFP}), because at least one of the parameters must be $O(1)$ to allow the left side to match its right side, which is $<-1$. This is consistent with the physical picture of the Ikeda instability, in which two adjacent cavity modes are driven by a pump field half-way between them, leading to a $2 t_R$ oscillation. These Ikeda instabilities are present on the positive slope branch of the characteristic, but because the driving is anti-resonant, the input field $F_{in}$ must be much larger than for mean-field instabilities.
\section{Conclusion}
In summary, we have, by adapting and extending a recently developed gain circle technique, obtained analytic dispersive and diffractive instability threshold formulae for Kerr Fabry-Perot cavities. We have thereby unified and greatly extended the results of four decades of research into counter-propagating fields in Kerr media. In the diffractive case, our gain circle approach is valid from arbitrarily-high finesse all the way to arbitrarily-low, or even zero, finesse. In the last limit our model correctly reproduces  SFM instability threshold formulae. In the dispersive case we neglected walk-off,  and obtained a dispersive instability threshold formula which holds for any practical Fabry-Perot cavity. Our formula agrees with recent results in the LLE limit, while generalizing them to arbitrary finesse, and also allows consideration of Ikeda instabilities. Because the gain-circle method is both general and well-suited to numerical implementations, it should prove widely applicable for the determination of key instabilities, leading to nonlinear modes and solitons, in photonics and quantum technology devices based on optical resonators. 

\section{Appendix}

The linearized governing equations (4) of the main text 
constitute a boundary-value problem after the time derivative has been resolved in either the diffractive or dispersive case. The two-point boundary conditions (5) in the main text 
suggest a solution based on the principle of superposition, and we choose two linearly-independent solutions, $(f_1(z),b_1(z))$ and $(f_2(z),b_2(z))$, that satisfy the boundary condition at $z=0$. The cavity boundary conditions (5) of the main text 
mean that there is only one free parameter in the initial conditions, and so the general solution of the perturbation equations (4) 
can be found as a linear superposition of any two linearly independent solutions. Since only numerical solutions may be available, construction of the locus of such output solutions is an effective method of finding a solution satisfying the other boundary condition, i.e. the locus of $f(-L) = (1-u)f_1(-L) + uf_2(-L)$ and $b(-L) = (1-u)b_1(-L) + ub_2(-L)$ on variation of  the real parameter $u \in [0,1]$. We recently showed (see [7] in the main text) 
that framing this problem in terms of the gain $g = f/b$ leads to intuitive results since the gain at the mirror ($z=-L$) traces out circles in the complex plane as the parameter $u$ is tuned.  

To demonstrate the existence of the gain circles we note that a little algebra shows that $g=f/b$ is given by a simple analytic formula in terms of $g_{1} =f_1/b_1$ and $g_{2} =f_2/b_2$,
\begin{equation}
\label{gaincircle}
g(u)=g_{2} +(g_{1} - g_{2})/(1+We^{i\phi_b}),
\end{equation}
where $W = u (1-u)^{-1} |(b_2/b_1)|$ is real, while $\phi_b$ is the phase of $b_2/b_1$. All parameters in (\ref{gaincircle}) are $z$ dependent, but explicit $z$-dependencies have been suppressed here, for clarity.

All possible gain values as $u$ is varied can thus be calculated from the variation of $(1+We^{i\phi_b})^{-1}$. This locus turns out to be a circle in the complex plane, with center $(1-e^{2i\phi_b})^{-1}$ and radius $|1-e^{2i\phi_b}|^{-1}$. Since a circle in the complex plane remains a circle when multiplied by any complex number and translated by any other, it follows that the locus of the gain function is also a circle, which we term the gain circle. Clearly $g(u)= g_{1},g_{2}$ for $u=0,1$, so both these values lie on the gain circle, as they must. The center $C_g$ of the gain circle lies at $g_{2} +(g_{1}- g_{2})/( 1-e^{2i\phi_b})$, while its radius $R_g$ is $|g_{1} - g_{2}|/| 1-e^{2i\phi_b}|$, which is a general result that applies to any gain-circle analysis. 

In order to match the boundary conditions at $z=-L$ we must set the gain $g = r e^{i \phi}$, where $r$ is the mirror reflectivity and $\phi$ is the total cavity phase. It is therefore convenient to express the points on the gain circle with center $C_g$ and radius $R_g$ in terms of polar coordinates $(r,\phi)$,
\begin{equation}
\label{genthresh}
r^2 - 2 r Re\left(C_g e^{-i \phi} \right) = R_g^2 - |C_g|^2.
\end{equation}
For this Fabry-Perot cavity problem, exact expressions for the center $C_g$ and radius $R_g$ of the gain circle can be determined
\begin{eqnarray*}
Re(C_g) &=& \frac{\cos \psi_1 \cos \psi_2 - (\psi_2 /\psi_1 ) \sin \psi_1 \sin \psi_2}{\cos \psi_1 \cos \psi_2  + (\psi_2 /\psi_1 ) \sin \psi_1 \sin \psi_2}, \\
Im(C_g) &=& \frac{(\theta/\psi_1) \sin \psi_1 \cos \psi_2  + (\psi_2 /\theta) \cos \psi_1 \sin \psi_2}{\cos \psi_1 \cos \psi_2  + (\psi_2 /\psi_1 ) \sin \psi_1 \sin \psi_2}, \\
R_g^2 &=& |C_g|^2 - 1.
\end{eqnarray*}
The last of these equations eliminates $R_g$ from the general threshold formula (\ref{genthresh}) above, leading to the simple and elegant form of (9) 
in the main text.

The real and imaginary parts of $C_g$ can be written in the alternative form
\begin{eqnarray*}
Re(C_g) &=& \frac{1 - \psi_2^2 \tanc \psi_1 \tanc \psi_2}
{1  +  \psi_2^2 \tanc \psi_1 \tanc \psi_2}, \\
Im(C_g) &=& \frac{\theta \tanc \psi_1  + (\psi_2^2 /\theta)  \tanc \psi_2}
{1  +  \psi_2^2 \tanc \psi_1 \tanc \psi_2}.
\end{eqnarray*}
where $\tanc x = \sin x/(x\cos x)$ is an even function of $x$ with $\tanc (0) =1$. Using this form,  $C_g$ is clearly given to second order by $C_g = 1 -2\psi_{2}^2 + i(\theta +\psi_{2}^2/\theta)$, which directly leads to mean-field threshold formula (\ref{HFFP}). Note that $\psi_{2}^2/\theta = \theta -2(1+G)IL$, and that dispersive instability is well approximated by setting $G=0$ in the above expressions for $C_g$.

\end{document}